\documentclass{article}

\usepackage{arxiv}

\usepackage[utf8]{inputenc} 
\usepackage[T1]{fontenc}    
\usepackage{hyperref}       
\usepackage{url}            
\usepackage{booktabs}       
\usepackage{amsfonts}       
\usepackage{nicefrac}       
\usepackage{microtype}      
\usepackage{lipsum}
\usepackage{graphicx}
\usepackage{amsmath}
\usepackage{algorithm,algorithmic}
\usepackage{multirow}
\usepackage{enumitem}
\graphicspath{ {./images/} }

\title{Enhancing spatial auditory attention decoding with neuroscience-inspired prototype training}

\author{
 Zelin Qiu \\
  Institute of Acoustics\\
  Chinese Academy of Sciences\\
  \texttt{qiuzelin@hccl.ioa.ac.cn} \\
   \And
 Jianjun Gu \\
Institute of Acoustics\\
Chinese Academy of Sciences\\
\texttt{gujianjun@hccl.ioa.ac.cn} \\
  \And
 Dingding Yao \\
Institute of Acoustics\\
Chinese Academy of Sciences\\
\texttt{yaodingding@hccl.ioa.ac.cn} \\
  \And
Junfeng Li \\
Institute of Acoustics\\
Chinese Academy of Sciences\\
\texttt{lijunfeng@hccl.ioa.ac.cn} \\
}

\begin{document}
\maketitle
\begin{abstract}
The spatial auditory attention decoding (Sp-AAD) technology aims to determine the direction of auditory attention in multi-talker scenarios via neural recordings. Despite the success of recent Sp-AAD algorithms, their performance is hindered by trial-specific features in EEG data. This study aims to improve decoding performance against these features. Studies in neuroscience indicate that spatial auditory attention can be reflected in the topological distribution of EEG energy across different frequency bands. This insight motivates us to propose Prototype Training, a neuroscience-inspired method for Sp-AAD. This method constructs prototypes with enhanced energy distribution representations and reduced trial-specific characteristics, enabling the model to better capture auditory attention features. To implement prototype training, an EEGWaveNet that employs the wavelet transform of EEG is further proposed. Detailed experiments indicate that the EEGWaveNet with prototype training outperforms other competitive models on various datasets, and the effectiveness of the proposed method is also validated. As a training method independent of model architecture, prototype training offers new insights into the field of Sp-AAD. 
\end{abstract}


\section{Introduction}
\label{sec:introduction}

Individuals with normal hearing can track the sounds of interest in complex acoustic environments, a phenomenon known as the ``cocktail party effect” \cite{cherry1953some}. However, those with hearing impairments often struggle with this task \cite{festen1990effects, peelle2016neural}, even with the aid of hearing devices \cite{bronkhorst2000cocktail}. This issue arises because current hearing devices typically reduce background noise indiscriminately, failing to amplify the sounds of interest to the user \cite{clark2014technology, green2022speech, saki2016automatic}. To address this issue, brain-controlled hearing aids have been proposed. These devices are designed to detect the user’s auditory attention and selectively amplify the speech of the target speaker while suppressing interfering speakers \cite{hjortkjaer2024real, aroudi2020cognitive, ceolini2020brain}. With the advancement of speech signal processing technology \cite{zheng2023sixty, wang2018supervised}, the success of  such hearing aids depends on the precise detection of the listener’s auditory attention via neural signals.

The technology of determining auditory attention through neural signals is known as auditory attention decoding (AAD). Given its potential for practical applications, there has been a notable shift towards employing scalp electroencephalogram (EEG) signals for AAD instead of intracranial EEG or magnetoencephalography (MEG) signals. Notably, O’Sullivan’s pioneering work demonstrated the feasibility of decoding auditory attention from single-trial EEG \cite{o2015attentional}, leading to substantial advancements in this domain \cite{puffay2023relating, geirnaert2021electroencephalography, su2022stanet, geirnaert2020fast, accou2023decoding, qiu23_interspeech}.
Currently, AAD techniques fall into two main categories \cite{geirnaert2021electroencephalography, rotaru2024we}: (1) stimulus reconstruction methods, which convert EEG into speech representations like speech envelope or Mel spectrum and identify the attended speech stream by comparing its similarity (e.g., Pearson's correlation) with the actual speech representation; and (2) direct classification methods, which employ classifiers to identify the direction of the sound source being attended to by the user (typically binary classification of left and right), also known as spatial auditory attention decoding (Sp-AAD). 

Research \cite{su2022stanet,rotaru2024we} has demonstrated two principal advantages of Sp-AAD over stimulus reconstruction methods. Firstly, Sp-AAD can directly decode the orientation of the target speech source from the user’s EEG signals, without relying on speech information. Secondly, it achieves precise decoding without requiring an extended decision window. These characteristics enhance the applicability of this decoding method in practical scenarios, thereby sparking a wave of innovative research in the domain. For instance, Vandecappelle et al. \cite{vandecappelle2021eeg} incorporated convolutional neural networks into Sp-AAD to better model the nonlinear characteristics of EEG signals; Cai et al. \cite{su2022stanet} introduced temporal and channel attention into Sp-AAD, allowing the decision system to focus more on signals from specific channels and time periods; Xu et al. \cite{xu2024densenet} employed 3D convolution to extensively explore the spatial distribution of EEG electrodes.

Recent studies, including works by Su et al. \cite{su2022stanet}, Pahuja et al. \cite{pahuja2023xanet}, and Xu et al. \cite{xu2024densenet}, have reported impressive accuracy in Sp-AAD. However, these studies have yet to fully account for the impact of trial-specific features on decoding performance, leaving their true efficacy unconfirmed. Influenced by factors such as the subject’s mental state, neural signals exhibit temporal autocorrelation within a certain time frame. For example, \cite{prent2020dynamics} successfully predicted differences in creativity across subjects by the autocorrelation of alpha band energy, indicating that EEG signals within a certain time window share similarities. As subjects begin to rest or switch tasks, the characteristics of EEG signals shift significantly, resulting in similar statistical properties within the same trial and greater differences between different trials \cite{rotaru2024we}, thereby generating so-called trial-specific fingerprints \cite{rotaru2024we} or block-level effects \cite{bao2023block}. 

Given that the statistical variances between trials may surpass those between categories \cite{bao2023block}, decoders tend to overfit to trial-specific features, resulting in markedly inconsistent performance under different data partitioning strategies  \cite{puffay2023relating, bao2023block, rotaru2024we}. On one hand, when both the test and training sets originate from the same trials, i.e., employing a within-trial data partitioning strategy, decoders can leverage trial-specific features to determine the origin of test data, thereby assigning accurate labels and achieving high decoding accuracy. This strategy is commonly used in many current studies \cite{su2022stanet, pahuja2023xanet, xu2024densenet}. On the other hand, when the test and training sets span different trials, i.e., employing a cross-trial data partitioning strategy, decoders lack access to trial-specific information for inference. Consequently, trial-specific features hinder the capture of auditory attention features for the decoder, resulting in a significant drop in decoding accuracy, which may even reach chance levels \cite{puffay2023relating}. While existing research has explored the influence of trial-specific features on EEG decoding \cite{puffay2023relating, rotaru2024we, li2020perils}, there remains a lack of focused investigation aimed at improving decoding accuracy against these features.

To address this issue, it is crucial to enhance auditory attention features for easier capture while diminishing the harmful trial-specific features. Previous research has shown a strong correlation between spatial auditory attention and the topological distribution of energy in specific EEG frequency bands, particularly the lateralization of alpha and gamma band energies \cite{frey2014selective, wostmann2018opposite, deng2020topographic}. Inspired by these neurological phenomena, we posit that the energy distributions in different EEG frequency bands are crucial for Sp-AAD. Drawing inspiration from this insight and the EEG analysis methods in neuroscience \cite{roach2008event} that enhance the signal-to-noise ratio (SNR) by averaging EEG signals across different trials, this study proposes a new method called Prototype Training. This method constructs prototype samples by averaging the energy of multiple samples with the same label, enabling the neural network to better capture the energy distribution features which is related to auditory attention. This neuroscience-inspired method offers two key benefits: (1) Each training sample is a combination of samples from the same directional category, preserving crucial energy distribution features and attenuating random non-directional noise through superposition, making the relevant features more pronounced. (2) Since each prototype sample contains information from different trials, the trial-specific characteristics in the training set are significantly reduced, thereby decreasing the model's tendency to fit them.

In the proposed method, prototype samples are constructed using a straightforward summation approach to enhance the energy distribution features. Considering that time-domain signals contain phase information, direct summation could lead to undesirable outcomes, such as signal cancellation due to opposite phases. Consequently, the frequency domain energy of EEG is considered as the input. Building on this, we further propose a simple-structured model named EEGWaveNet. This model uses the wavelet transform of EEG as input, aligning with the requirements of the proposed prototype training. Moreover, compared to time-domain signals, the time-frequency energy spectrum explicitly represents the energy distribution of EEG, facilitating the extraction of auditory attention features.

This study introduces a novel method in the field of Sp-AAD, and the contributions can be summarized as follows:
\begin{itemize}
	\setlength{\itemsep}{1pt}
	\setlength{\parsep}{1pt}
	\setlength{\parskip}{1pt}
	\item A neuroscience-inspired method called prototype training is proposed, which enhances auditory attention features and reduces the impact of trial-specific characteristics in the Sp-AAD task.
	\item Based on prototype training, a decoder named EEGWa-veNet is introduced. It uses the energy of EEG in time-frequency domain as input, enabling the model to better extract features related to auditory attention.
	\item The effectiveness of the proposed methods is validated through extensive comparative experiments and data visualization.
	
	\item The comprehensive results, involving the training of over 50,000 distinct models, provide a unified performance benchmark for various models with different data partitioning strategies, contributing to the Sp-AAD field.

\end{itemize}
The remainder of this paper is organized as follows: Section \ref{section2} provides a concise definition of the Sp-AAD problem. Section \ref{section3} elaborates on the prototype training approach and the EEGWaveNet. Section \ref{section4} details the experimental setups, including the datasets, comparative models, data partitioning strategies, preprocessing methods and training details. Section \ref{section5} presents the experimental results in depth. Section \ref{section6} offers some discussions, and finally, a brief summary is presented in Section \ref{section7}.

\begin{figure*}[h]

	\centering
	
	\includegraphics[width=\linewidth]{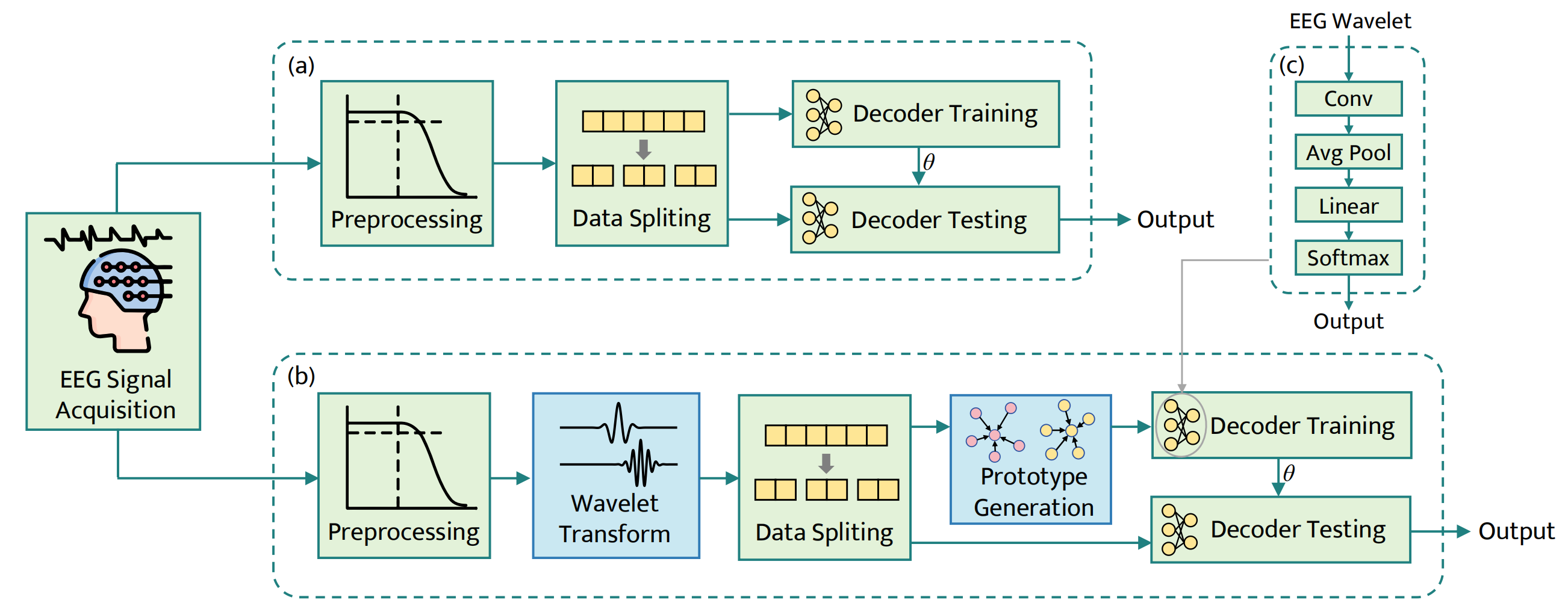}
	\caption{An overview of the proposed method. (a) The pipeline of existing mainstream methods. (b) The pipeline of the proposed method, incorporating time-frequency transform and prototype generation processes. (c) The model architecture of the EEGWaveNet.}
	\label{figure1}
\end{figure*}
\section{Problem Formulation}
\label{section2}
This study focuses on scenarios involving dual speakers. In such scenarios, a subject equipped with an EEG recording apparatus is presented with two distinct audio sources, one on each side (left and right). In each trial, the subject is instructed to focus on the sound source in a specific direction. The EEG data collected from the subject is denoted as $E_A$. After preprocessing, the EEG is segmented into parts of designated lengths, with the entire collection of segments denoted by $\mathcal{E}$. Each segment, referred to as a decision window, is denoted as $E \in \mathbb{R}^{T \times C}$, where $T$ represents the length of the EEG signal, and $ C $ indicates the number of EEG channels. The goal of Sp-AAD is to develop a decoder $\mathfrak{D}(\cdot) $ that takes $ E $ as input to determine the direction of the subject’s attention:
\begin{equation}
	p = \frak{D}(E_w) 	
\end{equation}
\begin{equation}
\hat{y}= \underset {j\in\{0, 1\}}  {\operatorname {arg\, max} }\, p[j]
\end{equation}
In this context, the output $\hat y$ assumes a binary form, with 0 denoting the left and 1 denoting the right. The decoding is considered correct if $\hat y$ matches the actual value $y$.

\section{Method}
\label{section3}
Prevailing Sp-AAD methods predominantly employ neural networks for end-to-end feature extraction and classification from time-domain EEG signals, as illustrated in Fig. \ref{figure1}(a). However, the pronounced nature of trial-specific features in EEG data can negatively impact the model’s ability to learn auditory attention features. Moreover, time-domain signals may not be optimal for Sp-AAD, potentially limiting the model’s capacity to extract these features. To address these issues, this study proposes a novel Sp-AAD pipeline, illustrated in Fig. \ref{figure1}(b), which incorporates a prototype generation module and a feature transformation module to enhance existing methodologies.

\subsection{Prototype Training}
Empirical studies reveal that spatial auditory attention elicits distinct  topological distributions of brain signal energy in specific frequency bands  \cite{frey2014selective, wostmann2018opposite, deng2020topographic}. For instance, alpha-band energy tends to increase in the cerebral hemisphere ipsilateral to the attended direction and decrease contralaterally. However, in single samples, such features could be less pronounced than the inherent trial classification characteristics, leading neural networks to focus on the latter. To mitigate this issue, a neuroscience-inspired method termed prototype training is introduced in this study. This method aims to enhance auditory attention features while
diminishing the trial classification attributes of the input signals. Unlike conventional prototype learning, which involves the network learning a fixed set of prototypes \cite{wang2019eeg, yang2018robust, biehl2016prototype}, the proposed approach generates diverse prototype training samples by merging multiple samples of the same class. Specifically, during training, several decision windows with the same label are randomly sampled from different trials for each input instance:
\begin{gather}
	\text{Sample:}\ (E_{w,1}, l_1), (E_{w,2}, l_2),...,(E_{w,K}, l_K)
\end{gather}
where $K$ is the number of decision windows used to generate a prototype sample and $E_w$ represents a decision window in the time-frequency domain. Each decision window shares the same label value, $l$, which is either 0 or 1, representing the target direction to the left or right, respectively. These decision windows are then combined with random weights to form a prototype sample:
\begin{gather}
	E_p=\displaystyle{ \sum_{i=1}^K} w_iE_{w,i}
\end{gather}
The prototype sample, denoted as $ E_p$, is formulated by employing positive weights $w_i $ that are randomly sampled under the constraint $ \displaystyle{\sum_{i=1}^K}w_i=1$. The generated prototype samples are subsequently used as input for neural network training. This approach offers the following potential benefits: (1)  Each prototype sample is a superposition of multiple homogeneous samples, which preserves auditory attention features while reducing non-directional random noise. (2) Each prototype sample contains information from multiple trials, effectively attenuating the trial-specific characteristics. (3) This method is only used during the training phase and does not introduce additional computational load during the model’s inference.

\subsection{Time-Frequency Representations}

To conduct the proposed prototype training, the commonly employed time-domain EEG is avoided due to phase issues. Instead, we consider the energy spectrum of EEG. Given the dynamic nature of brain processes during auditory perception, capturing the temporal dynamics of EEG is essential. Therefore, we propose that using the time-frequency spectrum of EEG as the input is beneficial. Compared to time-domain signals, the time-frequency spectrum offers a distinct advantage for Sp-AAD: The energy distribution of EEG is explicitly represented in the energy spectrum, facilitating the extraction of auditory attention features.

The continuous wavelet transform (CWT) is a prevalent technique in the time-frequency analysis of EEG signals. Compared to the short-time Fourier transform (STFT), CWT allows for adaptive lengths of analysis windows \cite{canal2010comparison, roach2008event}, achieving a dynamic balance between time resolution and frequency resolution. In light of this, this study proposes EEGWaveNet, which leverages the wavelet transform of EEG as input to better extract auditory attention features. First, CWT is applied to the preprocessed EEG $E_{proc}$:
\small
\begin{gather}
	\text{CWT}\{E_{proc}(t, i)\}=  \displaystyle{\int ^{\infty}_{-\infty}\frac{1}{\sqrt{\alpha}} E_{proc}(\tau, i) \psi \left(\frac{1}{\alpha}(\tau-t) \right ) d\tau}   \label{equation2} 	\\
	E_{TF} = \log {[\text{CWT}\{E_{proc}\} ]^2}  \label{equation3} 
\end{gather}
\normalsize

In Eq. \ref{equation2}, $i$ denotes the channel index, $\alpha$ is the scale factor of the wavelet in the time domain, and $\psi$ represents the Morlet wavelet \cite{morlet1982wave}. The logarithmic transformation is applied to compress the dynamic range of the EEG energy. Subsequent to the wavelet transform, the EEG data is segmented into decision windows of specified lengths. Each decision window is represented as $E_w\in \mathbb{R}^{C\times T_w\times F}$, where $T_w$ indicates the length of the signal, and $F$ corresponds to the number of frequency points.
\subsection{Decoder and Loss Function}
To demonstrate the effectiveness of the proposed method, we employ a simple convolutional neural network called EEGWaveNet as the decoder $ \mathfrak{D} $, as shown in Fig. \ref{figure1}(c). Initially, a two-dimensional convolutional layer \cite{yamashita2018convolutional} is utilized to capture the spatial temporal features of the EEG, followed by batch normalization \cite{ioffe2015batch} and a PReLU activation function \cite{he2015delving}:
\begin{gather}
	E_{p,c}=\text{PReLU}(\text{BN}(\text{Conv2d}(E_p)))
\end{gather}
Here, $E_{p,c} \in \mathbb{R}^{C_1 \times T_1 \times F_1} $ represents the output feature of the convolutional module.
Then, an averaging operation over the temporal dimension yields $E_{p,Avg} \in \mathbb{R}^{C_1 \times F_1} $. Subsequently, by flattening $E_{p,Avg} \in \mathbb{R}^{C_1 \times F_1} $ and passing it through a linear layer followed by a softmax layer, the final output probability is obtained:
\begin{gather}
	p=\text{Softmax}(\text{Linear}(E_{p,Avg}))
\end{gather}
The neural network is trained using a cross-entropy loss function:
\begin{gather}
	L= -[y \log p + (1-y)\log (1-p)]   
\end{gather}
For testing, the target direction is predicted using the original decision windows:
\begin{gather}
	p = \frak{D}(E_w)\\
	\hat{y}= \underset {j\in\{0, 1\}}  {\operatorname {arg\, max} }\, p[j]
\end{gather}
The proposed prototype training can diminish the influence of trial-specific features while enhancing the neural network’s ability to extract features related to auditory attention. To facilitate a clearer understanding, the training process is summarized in Algorithm \ref{algorithm1}.
\begin{algorithm}
	\caption{Training Algorithm for The Proposed Method}
	\renewcommand{\algorithmicrequire}{\textbf{Input:}} 
	\renewcommand{\algorithmicensure}{\textbf{Output:}} 
	\label{algorithm1}
	\begin{algorithmic}
		\REQUIRE EEG training set $\mathcal{E}_w$ and corresponding spatial attention label $\mathcal{L}$,  initial decoder $\frak{D}_{\theta}$, prototype sampling number $K$                
		\ENSURE optimal learned model
		\WHILE{$convergence \ condition \ is \  not \ satisfied$}
		\STATE Sample a batch of decision windows $\{   E_{w, 1\sim B}  \}\in \mathcal{E}_w$ and corresponding labels $\{l_{1\sim B}\}\in \mathcal{L}$; \\
		
		\FOR{$i=1$ to $B$}
		\STATE Sample $\{E_{w,i_1\sim i_{K-1}}\}$ that share the same label with $E_{w,i} $ from $\mathcal{E}_w$; \\
		\STATE Sample weights $\{w_{1\sim K}\}$ from $\text{Uniform}[0.1,1]$;
		\STATE Compute $w_{sum}=\displaystyle{\sum_{j=1}^{K}}w_j$;
		\FOR{$j=1$ to $K$}
		\STATE $w_i  \leftarrow w_i / w_{sum}$ 
		\ENDFOR
		\STATE $E_{p, i}  \leftarrow w_K  E_{w, i} + \displaystyle{\sum_{j=1}^{K-1}}w_j  E_{w,i_j}$ 
		
		\ENDFOR
		
		\STATE Calculate output $\{p_{1\sim B}\}$ with the decoder $\frak{D}_{\theta}$ and the input $\{   E_{p, i}, i=1, 2,...,B  \}$;
		\STATE Calculate the cross entropy loss with the output $\{y_{1\sim B}\}$ and the labels $\{l_{1\sim B}\}$;
		\STATE Update model parameters $\theta$ through back-propagation;
		
		\ENDWHILE

	\end{algorithmic}
\end{algorithm}

\section{Experimental Setups}
\label{section4}
\subsection{Data Specifications}
In this study, three commonly used EEG datasets for Sp-AAD, namely Das-2016 \cite{das2016effect}, Fuglsang-2018 \cite{fuglsang2017noise}, and Fuglsang-2020 \cite{fuglsang2020effects}, were employed for evaluation. In these datasets, subjects were exposed to simultaneous auditory stimuli from both left and right directions and were instructed to focus on one specific direction in each trial. For detailed information about the datasets, refer to \cite{das2016effect, fuglsang2017noise,fuglsang2020effects}.

For Das-2016 and Fuglsang-2018, all data were used. For Fuglsang-2020, only data from subjects with normal hearing in a cocktail party scene (a total of 32 trials per subject) were used. One subject experienced an interruption during data recording (Subject 24), thus, we only retained data from the remaining 21 subjects. Additionally, the first six seconds of each trial were discarded to avoid initial target voice-only segments. For clarity, the datasets information is summarized in Table \ref{table1}, noting that this summary only accounts for the data used in this study.
\begin{table*}[]
	\fontfamily{ptm}\selectfont
	\centering
	\scriptsize

	\tabcolsep=0.15cm
	\renewcommand{\arraystretch}{1.2}
	\caption{Details of the three EEG datasets used in the experiments.}
	\label{table1}
	\begin{tabular}{@{}c|c|c|c|c|c@{}}
		\toprule
		Dataset       & Subjects & Language & Spatial Locus of The Stimuli & Trials Per Subject & Duration Per Subject                                      \\ \midrule
		Das-2016 \cite{das2016effect}     & 16       & Dutch  & +90°/-90°                    & 20                 & 72 min (8 trials $\times$ 6 min/trial + 12 trials $\times$ 2 min/trial) \\
		Fuglsang-2018 \cite{fuglsang2017noise} & 18       & Danish   & +60°/-60°                    & 60                 & 50 min (60 trials $\times$ 50 s/trial)                           \\
		Fuglsang-2020 \cite{fuglsang2020effects} & 21       & Danish   & +90°/-90°                    & 32                 & $\sim$23.5 min(32 trials $\times$ 44 s/trial)                    \\ \bottomrule
	\end{tabular}
\end{table*}

\subsection{EEG Preprocessing}

This study implemented uniform data preprocessing protocols across the three datasets. To validate the model’s true performance and avoid potential biases, similar to the approach in \cite{rotaru2024we}, data normalization or artifact removal were not employed. Specifically, EEG data were first downsampled to 128 Hz after anti-aliasing filtering, then a Butterworth band-pass filter ranging from 1 to 50 Hz was applied. Finally, the EEG were re-referenced to the average.

\subsection{Contrastive Models}
To provide a comprehensive comparison, the following three models, which are representative in this field, were also evaluated:
\begin{itemize}
	\setlength{\itemsep}{1pt}
	\setlength{\parsep}{1pt}
	\setlength{\parskip}{1pt}
	\item \textbf{FB-CSP} \cite{geirnaert2020fast}: The filterbank common spatial pattern (FB-CSP) method is a traditional machine learning algorithm. It employs CSP filters \cite{blankertz2007optimizing} to process EEG signals across different frequency bands to better capture the spatial features of EEG, and then classifies the features using linear discriminant analysis (LDA) \cite{xanthopoulos2013linear}. This method represents the best non-neural network algorithm in the Sp-AAD domain. 
	\item \textbf{CNN} \cite{vandecappelle2021eeg}: The CNN method is one of the earliest neural network approaches proposed in the Sp-AAD domain. It primarily consists of a convolutional layer, two fully connected layers, and several activation functions. This method has been used as a baseline in many studies and represents a lightweight neural network approach in the Sp-AAD field. With a single two-dimensional convolution serving as its backbone, this method allows for an ablation comparison with the proposed EEGWaveNet. 
	\item \textbf{DenseNet-3D} \cite{xu2024densenet}: DenseNet-3D is a recently proposed neural network based method that fully considers the topological relationships between EEG channels and introduces three-dimensional convolution to model spatial and temporal information simultaneously. Besides, dense connections are employed to aid network training and prevent feature loss. This method achieves the state-of-the-art (SOTA) performance in Sp-AAD.
	
\end{itemize}

For FB-CSP and DenseNet-3D, the experiments were conducted based on the original code \footnote{https://github.com/exporl/spatial-focus-of-attention-csp.}$^,$\footnote{https://github.com/xuxiran/ASAD\_DenseNet.} provided by the authors, and for CNN, the original code\footnote{https://github.com/exporl/locus-of-auditory-attention-cnn.} was ported to the Python platform for the experiments.
\subsection{Data Partitioning Strategies}

\begin{figure}[h]
	
	\centering
	\includegraphics[width=150mm]{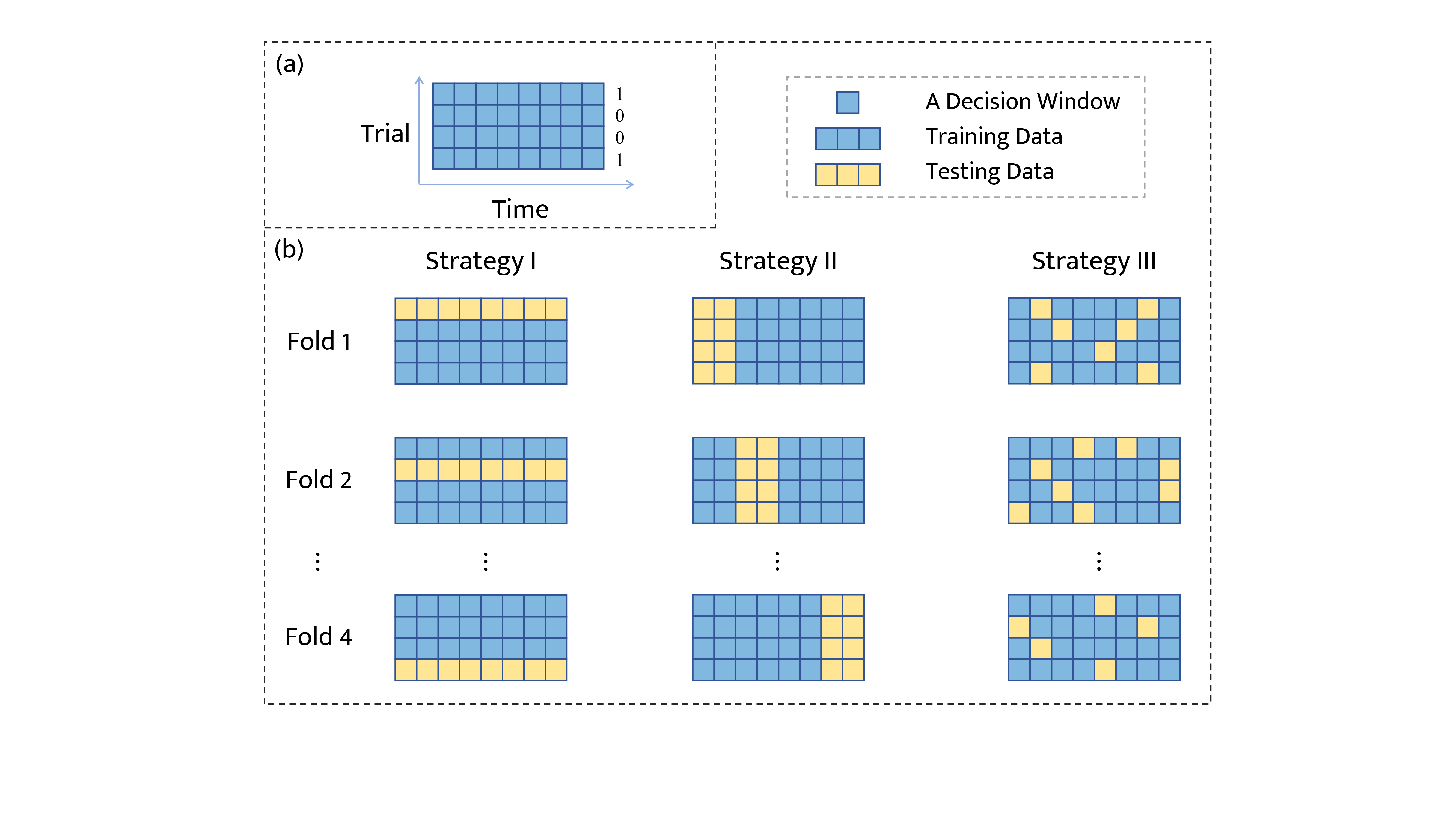}
	\caption{(a) A schematic diagram of data arrangement. (b) Schematic illustrations of three different data partitioning strategies under 4-fold cross-validation.}
	\label{figure2}
\end{figure}

To explore the impact of trial-specific features on decoding performance, three distinct data partitioning strategies were established: one cross-trial and two within-trial. For each strategy, a 4-fold cross-validation was employed to assess model performance. The three strategies utilized are depicted in Fig. \ref{figure2}, with details provided below:
\begin{itemize}
	\setlength{\itemsep}{1pt}
	\setlength{\parsep}{1pt}
	\setlength{\parskip}{1pt}
	\item \textbf{Strategy I \cite{qiu23_interspeech}}: In a given fold, one-quarter of the trials are randomly selected as test data, while the remaining trials are used as training data.

	\item \textbf{Strategy II  \cite{li2022esaa}}: Each trial is equally divided into four segments. In each fold, one segment from all trials are aggregated to serve as test data, with the remaining data used for training.

	\item \textbf{Strategy III \cite{su2022stanet}}: All trials are first divided into several decision windows based on the required window length. In each fold, one-quarter of all decision windows are randomly chosen as test data, with the remaining decision windows serving as training data. 
	
\end{itemize}
In the implementation of Strategy II, an overlap between training and test datasets is observed when the step size of the sliding window is less than the window length. To avoid data leakage, any decision windows that intersected with the test dataset were removed from the training set. In addition, to avoid random biases that may arise from data division, all experiments were repeated four times, with the final results being the mean accuracy over 16 folds.

\subsection{Training Details}
In this study, CWT was conducted using the FieldTrip toolbox \cite{oostenveld2011fieldtrip}. The sampling number per second was set to 10, and the frequency band ranged from 2 Hz to 50 Hz. For EEGWaveNet, a convolutional layer with nine filters, a kernel size of 3$\times$3, and a stride of 1$\times$1 was employed. 
All models were trained in a subject-dependent manner. Each neural network-based model underwent training of 50 epochs, utilizing a learning rate of 0.001 and a batch size of 16. For the proposed method, prototype samples matching the sample quantity of the original training dataset were generated in each epoch.

\section{Results and Analysis}
\label{section5}

\begin{table*}[h]
	\fontfamily{ptm}\selectfont
	\centering
	\scriptsize
	\tabcolsep=0.1cm
	\renewcommand{\arraystretch}{1.4}
	\caption{Decoding accuracy (\%) of various methods on different datasets with an one-second window length. Each result is obtained after averaging at the subject level. For each strategy, the best result of the five methods is highlighted in bold.}
	\label{table2}
	\begin{tabular}{@{}c|c|ccc|ccc|ccc@{}}
		\toprule
		& Dataset        & \multicolumn{3}{c|}{Das-2016} & \multicolumn{3}{c|}{Fuglsang-2018} & \multicolumn{3}{c}{Fuglsang-2020} \\ \midrule
		& Strategy       & I        & II       & III     & I          & II        & III       & I         & II        & III       \\ \midrule
		\multirow{5}{*}{     Model} & FB-CSP         & 69.43$\pm$15.77    & 80.39$\pm$10.10    & 83.06$\pm$8.85   & 53.62$\pm$5.72      & 61.25$\pm$5.55     & 64.23$\pm$4.90     & 66.75$\pm$19.09     & 76.39$\pm$14.74     & 79.28$\pm$13.05     \\
		& CNN            & 66.17$\pm$13.38    & 74.70$\pm$9.13    & 76.64$\pm$8.61   & 53.75$\pm$4.15      & 56.14$\pm$4.94     & 56.43$\pm$4.90     & 64.51$\pm$12.32     & 67.84$\pm$11.82     & 67.89$\pm$11.58     \\
		& DenseNet-3D    & 71.26$\pm$13.02    & 83.24$\pm$8.01    & 86.68$\pm$6.58   & 54.87$\pm$5.16      & 60.88$\pm$7.62     & 61.81$\pm$7.53     & 66.26$\pm$16.09     & 74.56$\pm$13.47     & 75.15$\pm$14.07     \\
		& EEGWaveNet-K1  & 72.71$\pm$13.27    & \textbf{85.15}$\pm$7.09    & \textbf{88.19}$\pm$6.06   & 53.76$\pm$4.95      & \textbf{63.83}$\pm$7.12     & \textbf{64.27}$\pm$7.36     & 68.73$\pm$17.66     & \textbf{82.18}$\pm$12.30     & \textbf{81.92}$\pm$12.32    \\
		& EEGWaveNet-K10 & \textbf{74.04}$\pm$14.36    & 83.40$\pm$7.95    & 85.33$\pm$6.58   & \textbf{55.41}$\pm$5.97      & 62.71$\pm$6.32     & 63.16$\pm$6.39     & \textbf{69.97}$\pm$17.14     & 79.72$\pm$12.20     & 79.95$\pm$11.82     \\ \bottomrule
	\end{tabular}
\end{table*}

\subsection{Main Results}
\label{section5_1}
We began our study by evaluating the performance of different models under various data partitioning strategies. Specifically, all three datasets were used to compare the performance of FB-CSP, CNN, DenseNet-3D, EEGWaveNet-K1, and EEGWaveNet-K10 under the three data partitioning strategies. Here, EEGWaveNet-K1 and EEGWaveNet-K10 refer to prototype training with $K=1$ and $K=10$, respectively, with EEGWaveNet-K1 equivalent to training on the original data without prototype training. The experiments were conducted using a 1.0-second window, which is commonly used in Sp-AAD, and the results are presented in Table \ref{table2}. The $Wilcoxon$ signed-rank test \cite{wilcoxon1992individual} was employed to compare the performance of different methods.

EEGWaveNet-K10 demonstrates superior performance compared to competing models under Strategy I. With the same architecture, EEGWaveNet-K10 consistently outperforms EEGWaveNet-K1 across the three datasets ($p<0.05$, $p<0.005$, $p<0.05$ ), supporting the hypothesis that prototype training can enhance auditory attention features, and thereby boost model performance in cross-trial scenarios. Moreover, EEGWaveNet-K10 outperforms the previously leading model, DenseNet-3D, by margins of 1.45\% ($p<0.05$), 0.54\%, and 3.71\% ($p<0.01$) in accuracy. Remarkably, EEGWaveNet’s parameter count (5.5k) is less than one percent of DenseNet-3D’s (754.8k), highlighting the efficiency of the proposed methodology. Compared to the CNN model, which also relies on a single convolutional layer, EEGWaveNet-K1 achieves improvements of 6.54\% ($p<0.005$), 0.01\%, and 4.22\% ($p<0.05$) across the three datasets, indicating that the time-frequency domain EEG is more effective for the model in identifying features pertinent to spatial auditory attention.

In contrast, when implementing within-trial Strategies II and III, EEGWaveNet-K10 performs worse than EEGWav-Net-K1 across all three datasets (for Strategy II: $p<0.005, p=0.085, p<0.001$; for Strategy III: $p<0.001, p=0.058, p<0.005$). This observation suggests that the proposed prototype training strategy effectively masks trial-specific features, enabling the model to focus more on auditory attention related features. Importantly, this finding also indicates that a model’s performance under within-trial strategies does not necessarily predict its performance under cross-trial conditions. Therefore, in Sp-AAD research, cross-trial strategies similar to Strategy I should be prioritized for a more accurate assessment of model capabilities..

Another notable finding is that, across all methods, the decoding accuracy under Strategy II significantly exceeds ($p<0.001$) that of Strategy I. Specifically, on Das-2016, the decoding accuracies of the five different methods under Strategy II are respectively 10.95\%, 8.53\%, 11.98\%, 12.44\%, and 9.36\% higher than those under Strategy I. On Fuglsang-2018, the increases are 7.63\%, 2.39\%, 6.00\%, 10.08\%, and 7.30\%; and on Fuglsang-2020, the improvements are 9.64\%, 3.32\%, 8.30\%, 13.44\%, and 9.75\%, respectively. Given the longer trial durations in Das-2016 (6 minutes or 2 minutes), this implies that even when trials are segmented into longer intervals, segments from the same trial still retain similar trial-specific characteristics. Additionally, while not consistently significant, decoding accuracy with Strategy III tends to surpass that of Strategy II across all datasets. This outcome is in line with prediction, as Strategy III involves dividing data from the same trial into shorter decision windows and randomly distributing them between the training and test sets. Consequently, the training set comprises numerous decision windows closely related to those in the test set, sharing more similar features and thereby leading to enhanced decoding performance. Therefore, we reemphasize the importance of employing cross-trial data division strategies. 

The results also reveal considerable variability in decoding accuracy across different datasets. Specifically, in cross-trial scenarios, the methods typically achieve a decoding accuracy of approximately 70\% on Das-2016, about 67\% on Fuglsang-2020, and merely 53\% on Fuglsang-2018, which is nearly at the level of random chance. This is consistent with the findings reported by \cite{geirnaert2021electroencephalography}. The diminished decoding accuracy observed in Fuglsang-2018 may be attributed to several factors: (1) The sound source locations in Fuglsang-2018 are at ±60°, offering less spatial separation compared to the other datasets, thus posing a greater challenge for the subjects. (2) The auditory stimuli presented to the subjects encompass simulated reverberation and background noise, which could potentially distract the subjects’ attention. (3) the speech materials in Fuglsang-2018 contain extended silent intervals compared to other datasets, during which participants’ focus may be disrupted by non-targeted speech. Therefore, it is suggested that the spatial attention effect in Fuglsang-2018 is suboptimal, and using it as a benchmark for Sp-AAD may not accurately reflect the models’ performance.

\begin{figure*}[h]
	
	\centering
	\includegraphics[width=\linewidth]{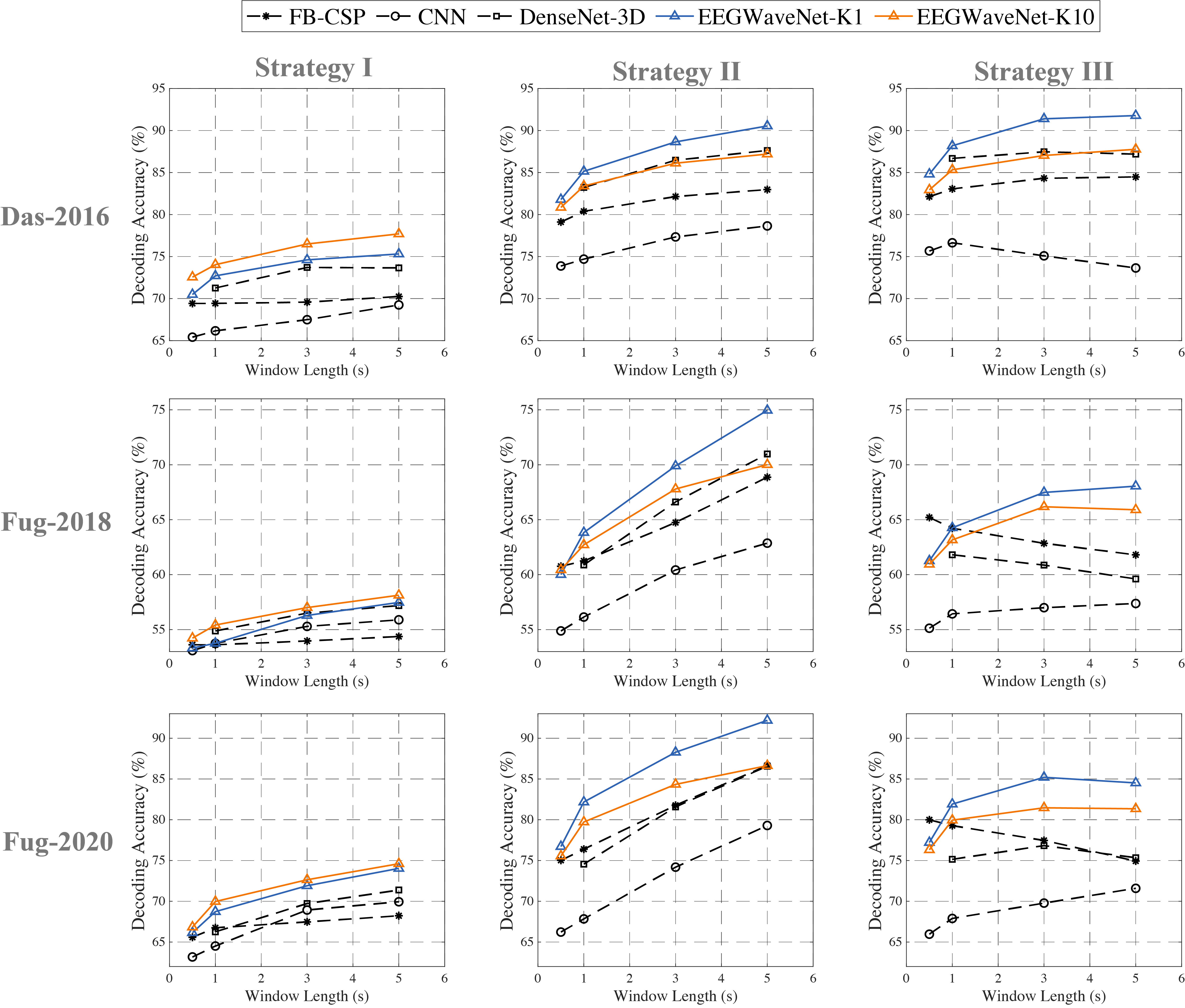}
	
	\caption{Performance of different models across varying window lengths. To clearly illustrate the datasets and data partitioning strategies used, nine charts are arranged in a matrix format. Horizontally, from left to right, they represent three data partitioning strategies: Strategy I, Strategy II, and Strategy III. Vertically, from top to bottom, they correspond to three datasets: Das-2016, Fuglsang-2018, and Fuglsang-2020. Each data point represents the average decoding accuracy of all subjects in the corresponding dataset. }
	\label{figure3}
\end{figure*}
\begin{figure*}[h]
	
	\centering
	\includegraphics[width=\linewidth]{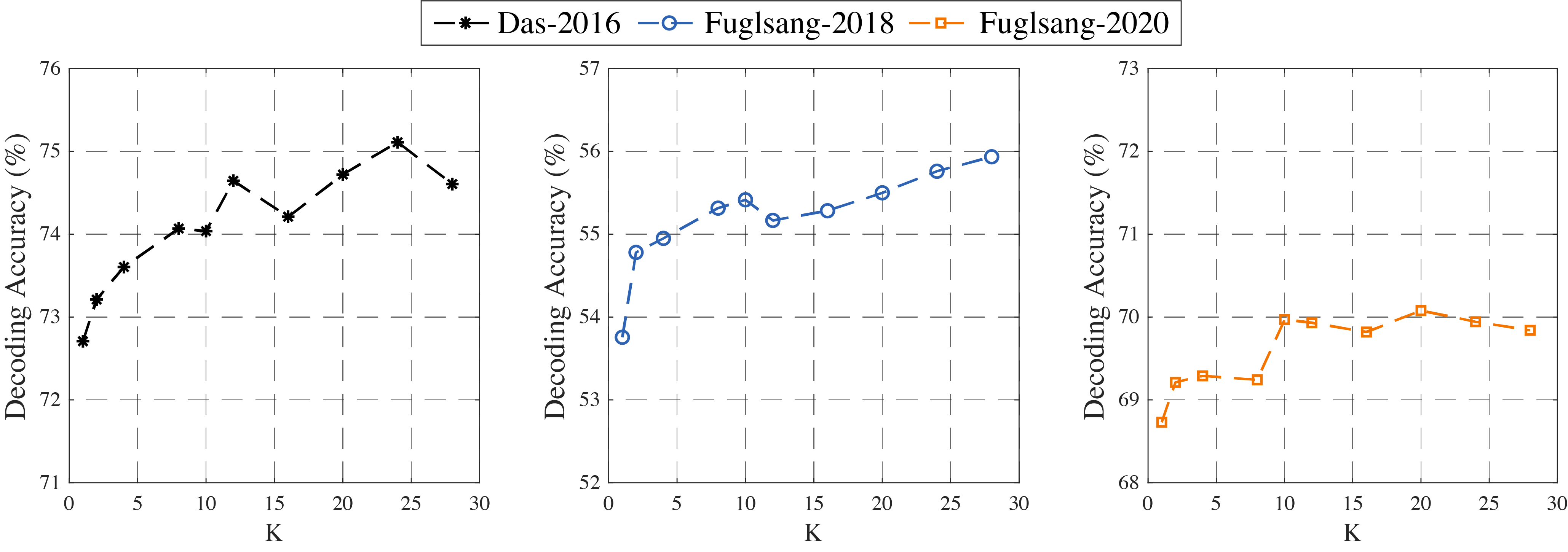}
	\caption{Illustration of decoding results obtained with different values of parameter $K$, representing, from left to right, the results on the Das-2016, Fuglsang-2018, and Fuglsang-2020 datasets respectively.}
	\label{figure4}
\end{figure*}
\subsection{Effect of Window Length}

The investigation extended to the analysis of decoding performance across various window lengths. Specifically, four window durations were examined: 0.5s, 1.0s, 3.0s, and 5.0s. To ensure a uniform number of decision windows across different window lengths, a consistent stride of 1.0s was employed, in alignment with Section \ref{section5_1}. However, there are two exceptions: (1) For the 0.5s window length, a different stride of 0.5s was adopted to utilize all available data. (2) For Strategy III, to prevent overlap between training and test data, a stride equivalent to the window length was adopted when the initial stride was shorter than the window duration, thereby avoiding an excessively small training dataset. The evaluation continued with the five models: FB-CSP, CNN, DenseNet-3D, EEGWaveNet-K1, and EEGWaveNet-K10. DenseNet-3D is incompatible with a 0.5s window duration, resulting in the absence of results for this specific window. The experimental results are illustrated in Fig. \ref{figure3}.

In cross-trial scenarios (Strategy I), EEGWaveNet-K10 demonstrates superior performance across various datasets and window lengths. On Das-2016, EEGWaveNet-K10 attains accuracies of 72.55\%, 74.04\%, 76.50\%, and 77.70\% for the respective window durations. On Fuglsang-2018, the accuracies are 54.22\%, 55.41\%, 57.00\%, and 58.13\%; and on Fuglsang-2020, the accuracies reach 66.84\%, 69.97\%, 72.64\%, and 74.61\%. These results surpass those of EEGWa\ -veNet-K1 and the other models under identical conditions, futher demonstrating that prototype training can enhance auditory attention features. Conversely, under Strategy II, except for the case of \textlangle Fuglsang-2018, 0.5s\textrangle, the performance of EEGWaveNet-K10 does not meet that of EEGWaveNet-K1. This observation is consistent with previous conclusion that prototype training can effectively obscure trial-specific characteristics, thus reducing their impact on decoding auditory attention.

In Fig. \ref{figure3}, the scale of the vertical axis is consistent across the three plots of the same dataset. The results clearly show that the decoding accuracy with Strategy II surpasses that with Strategy I for all datasets examined. Furthermore,  while an increase in decoding accuracy is observed as the window length extends under both strategies, the increase is more pronounced under Strategy II than under Strategy I. Specifically, a linear function was fitted to the decoding accuracy as a function of varying window lengths. Under Strategy I, the mean slopes for the five methods across the three datasets are 0.73, 0.64, and 1.30, respectively. In contrast, when employing Strategy II, the average slopes are significantly higher at 1.21, 2.28, and 2.79, representing increases of 65.8\%, 256.3\%, and 114.6\% over Strategy I. This indicates that trial-specific information plays a substantial role in ``enhancing'' decoding accuracy in non-cross-trial scenarios. As the window length is extended, the trial-specific features within the decision window assist the model in making more precise predictions. Furthermore, under Strategy II, the slopes for EEGWaveNet-K10 are 1.30, 2.10, and 2.27, all lower than those of EEGWaveNet-K1, which are 1.79, 3.16, and 3.18, further demonstrating that prototype training can mitigate the impact of trial-specific features.

When Strategy III is applied, there is a decline in the slopes for all models across different datasets compared to Strategy II, with some models even exhibiting negative slopes, such as \textlangle Das-2016, CNN\textrangle$\ $and \textlangle Fuglsang-2018, FB-CSP/DenseNet-3D\textrangle. This phenomenon can be attributed to the inverse correlation between window length and the number of training samples in Strategy III, which may hinder the model’s ability to learn features effectively.
\begin{figure*}[h]
	
	\centering
	\includegraphics[width=\linewidth]{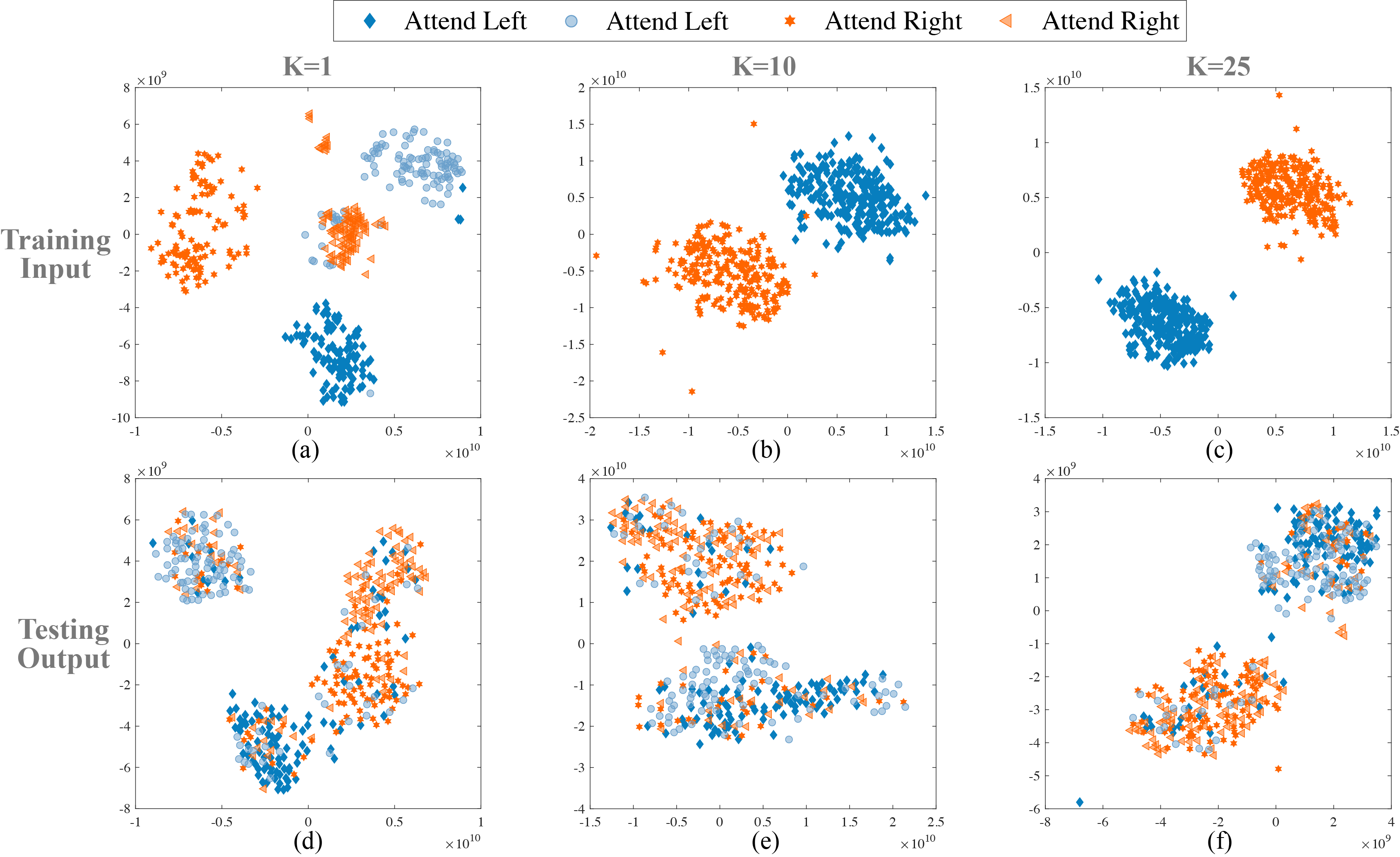}
	\caption{T-SNE visualization of input decision windows and output embeddings. (a) Projection results of decision windows from four random trials. (b)-(c) Projection results of prototype samples generated with $K=10$ and $K=25$, respectively. (d)-(f) 
		Projection results of high-dimensional embeddings produced by the trained EEGWaveNet with $K=1$, $K=10$, and $K=25$ using data from the the same four trials in the testset.
	}
	\label{figure5}
\end{figure*}

\subsection{Effect of The Sampleing Number K}

The influence of the prototype sampling number $K$ on the decoding accuracy was also investigated. For each dataset, experiments were conducted with $K$ set to 1, 2, 4, 8, 10, 12, 16, 20, 24, 28 respectively. Given that Strategy I offers a more accurate evaluation of model performance, it was utilized in this analysis. Fig. \ref{figure4} illustrates the decoding results for the three datasets with a window length of 1.0s, where each data point represents the mean accuracy across all subjects within the dataset.

The results on the Das-2016 and Fuglsang-2018 datasets show a consistent pattern: a notable initial rise in decoding accuracy as $K$ increases. In contrast, the Fuglsang-2020 dataset displays a less pronounced trend, with decoding accuracy experiencing a modest initial increase followed by fluctuations. Nonetheless, the lowest decoding accuracy across all datasets is observed when $K=1$, validating the efficacy of the proposed prototype training approach. Overall, a higher decoding outcome is achieved for all three datasets when $K$ is around 25. Besides, it is noteworthy that the prototype training method proposed is not confined to the EEGWaveNet model alone but is applicable to a broader range of models leveraging time-frequency domain features, offering a novel optimization technique to the Sp-AAD field.

\subsection{Data Visualization}

To further validate the effectiveness of the proposed prototype training, data visualization was employed. The impact of the proposed method on input data was first explored. Specifically, we utilized the t-SNE algorithm \cite{van2008visualizing} to project a subset of training data from a subject in Das-2016 onto a two-dimensional plane. When $K=1$, two trials from each direction were randomly selected, with their projection results depicted in Fig. \ref{figure5}(a). For $K=10$ and $K=25$, where each sample corresponds to multiple trials, an equivalent number of prototype samples to those in Fig. \ref{figure5}(a) were randomly selected, with the outcomes presented in Fig. \ref{figure5}(b) and \ref{figure5}(c).

Clearly, when $K=1$ (i.e., without prototype training), data from different trials cluster separately, and data with identical spatial labels do not form distinct clusters. This observation aligns with the description in the Introduction, indicating that EEG data demonstrate robust trial-specific classification traits, which may affect the model’s learning of spatial auditory attention. When $K=10$, the generated prototype samples cluster distinctly by spatial category, fulfilling our initial objective of enabling the model to better capture spatial attention features. Furthermore, when $K=25$, the clusters formed by the samples are more compact, corresponding to the higher decoding accuracy observed in Fig. \ref{figure4}.

Subsequently, the impact of the proposed method on output data was also explored. Specifically, we processed the same four trials’ data using trained EEGWaveNet with the same three $K$ values: 1, 10 and 25. The resulting high-dimensional embeddings were then projected onto a two-dimensional plane, as shown in Fig. \ref{figure5}(d)-(f). When $K=1$, although samples from the two trials with a rightward target direction (orange) cluster together, the samples still distinctly group by trial, indicating that the model has learned trial-specific features. However, when $K=10$, the output embeddings cluster distinctly by spatial category, and data from the same direction but different trials do not separate significantly. This suggests that the proposed strategy can mitigate the model’s reliance on trial-specific features, allowing it to focus more on spatial attention features.

\section{Discussion}
\label{section6}
\subsection{A Comprehensive Benchmark}

Many current studies on Sp-AAD utilize approaches similar to Strategy II for performance validation. Therefore, the high accuracy reported may be a result of overfitting to trial-specific features rather than genuinely decoding spatial auditory attention. While some works have highlighted this issue \cite{puffay2023relating, rotaru2024we}, a comprehensive exploration of how different data partitioning methods impact decoding accuracy is still lacking. Researchers lack precise information on the performance of common models under different data partitioning strategies.

This study aims to fill this gap. In addition to the proposed method, three models were selected for experimentation: the traditional FB-CSP algorithm, a simple-structured CNN, and the recently proposed high-performing DenseNet-3D. Specifically, separate models were trained under the following conditions:
\begin{itemize}[itemsep=0pt]
	\item 55 subjects across 3 datasets
	
	\item 3 data partitioning strategies
	\item 4 window lengths
	\item 5 models
	\item 4 repetitions, each with 4-fold cross-validation
\end{itemize}
Approximately 50,000 models were trained. The comprehensive experimental results provide benchmark performance for various models under different data partitioning strategies in the Sp-AAD domain, contributing to Sp-AAD research.
\subsection{Data-Driven vs Handcrafted Features}
Traditional methodologies for Sp-AAD typically encompass a two-stage process: initial handcrafted feature extraction followed by subsequent classification. However, the suitability of handcrafted features for the Sp-AAD task has been called into question \cite{su2022stanet}, prompting a shift towards the adoption of neural network-based end-to-end algorithms. These networks are trained to autonomously perform both feature extraction and classification, offering substantial benefits. Nevertheless, due to the overly simplistic labels of the Sp-AAD task—where all data within a trial share the same label—and the limited data per subject, models are prone to learning irrelevant information (termed mis-learning here), such as trial-specific features. The prevailing experimental paradigm presents challenges in avoiding this issue, as training data invariably comprises multiple samples from the same trial, potentially affecting the model’s performance.

This paper posits that although mis-learning cannot be completely eliminated, it can be reduced by guiding neural networks towards more effective extraction of features related to spatial auditory attention through some simple artificial designs. The EEGWaveNet model introduced in this paper adopts this approach. By applying wavelet transform, the EEG signals are segregated into distinct frequency bands, clearly exhibiting the energy distribution. This method, compared to pure time-domain approaches, markedly improves decoding accuracy. On the other hand, considering the complexity and individual variability of brain activity \cite{van2008individual}, it is crucial for models to autonomously learn appropriate features from data. Therefore, the model’s input should not be excessively manipulated, as it may otherwise lead to decreased performance. In summary, we advocate for a balanced approach: emphasizing auditory attention features through appropriate manual design while ensuring the network can autonomously extract critical information. The proposed EEGWaveNet satisfies these requirements. It employs the wavelet-transformed EEG as input, preserving sufficient information while presenting energy distribution in EEG directly, which is crucial for Sp-AAD.

\section{Conclusion}
\label{section7}
This paper introduces Prototype Training, a neuroscience-inspired method for Sp-AAD. This approach constructs prototypes from multiple trials to enhance energy distribution features related to auditory attention while diminishing trial-specific features. Based on prototype training, we further propose a decoder named EEGWaveNet, which employs the wavelet transform of EEG to better capture their energy distribution. Experimental results indicate that prototype training can improve model performance in cross-trial scenarios, and the effectiveness of EEGWaveNet is also demonstrated. Our results provide a comprehensive benchmark for different models under various data partitioning methods, a resource currently lacking in the field of Sp-AAD.

\bibliographystyle{unsrt}  
\bibliography{arxiv}  

\end{document}